# LONG-TERM VARIATIONS OF SOLAR MAGNETIC FIELDS DERIVED FROM GEOMAGNETIC DATA

K.Georgieva [1], B.Kirov [1], Yu.A.Nagovitsyn [2]

*1 – Space Research and Technologies Institute, Bulgarian Academy of Sciences, Sofia,*
*2 – Pulkovo Observatory of Russian Academy of Sciences, Pulkovo*

## 1. Introduction

Sunspots are dark spots on the solar surface associated with strong magnetic fields [Hale, 1908]. The magnetic field greatly reduces the convective transport of heat from below so the plasma in the magnetic flux tube piercing the solar surface is colder and appears darker. Therefore, it can be supposed that the number, area, and brightness of sunspots reflect the intensity of the solar magnetic fields.

However, contradicting results were published on the solar cycle and longer term variations of the magnetic fields in sunspots. Pevtsov et al. [2012] employed historic synoptic data sets from seven observatories in the former USSR covering the period 1957 – 2011 and found that the sunspot field strengths vary cyclically reaching maxima around sunspot maxima and minima around sunspot minima, with no indication of a secular trend in the last five sunspot maxima (cycles 19-23). In contrast, Penn and Livingston [2009, 2010], using the Zeeman-split 1564.8 nm Fe I spectral line at the NSO Kitt Peak McMath-Pierce telescope, found that the magnetic field strength in sunspots has been decreasing in time since 1990's, but with no dependence on the solar cycle. The explanation of this contradiction was offered by Nagovitsyn et al. [2012]: while Pevtsov et al. [2012] used only the biggest sunspots for the analysis, Penn and Livingston [2009, 2010] used all visible sunspots. During the period of 1998–2011, the number of large sunspots whose magnetic fields do show sunspot cycle variations but no long-term trend, gradually decreased, while the number of small sunspots whose weaker magnetic fields do decrease in time but have no sunspot cycle dependence, steadily increased.

The dependence of the sunspot's magnetic field on its area was first found by Houtgast and van Sluiters [1948] who also noted that the correlation varies in the course of the solar cycle. Ringnes and Jensen [1960], on the other hand, found no systematic changes of the correlation coefficient with epoch in the solar cycle, but very pronounced secular variations, opposite for sunspots of different size. An estimation of the sunspot magnetic fields and their variations can also be deduced from measurements of sunspots' brightness: a correlation exists between the sunspot's maximum brightness and its magnetic field – the darker the spot, the more intense its magnetic field [Abdussamatov, 1973]. A solar cycle dependence of the sunspot brightness was found by Albregtsen and Maltby [1978], with the emitted intensity from sunspots increasing from the beginning toward the end of the cycle. This implies that the sunspot magnetic fields decrease from the beginning to the end of the sunspot cycle, while Pevtsov et al. [2012] found that the magnetic fields increase and decrease with the number of sunspots.

As the causes and the time profiles of the variations in the correlations between the sunspot magnetic fields and the number, area, and brightness of sunspots are not quite clear, the sunspot data alone cannot be used as a proxy for deriving the variations of the sunspot magnetic fields for periods for no instrumental measurements are available. But the Earth is a sort of a probe reacting to interplanetary disturbances which are manifestation of the solar magnetic fields, so records of the geomagnetic activity can be used as diagnostic tools for reconstructing past solar magnetic fields evolution. In the present study we combine sunspot and geomagnetic data to estimate the long-term variations of sunspot magnetic fields.

## 2. Components of the geomagnetic activity and their long term variations

Geomagnetic disturbances are caused by solar activity agents carried by the solar wind. Feynman [1982] noticed that if a geomagnetic activity index (e.g. *aa*-index) is plotted as a function of the number of sunspots *R*, practically all points lie above a line (Fig.1) which represents the minimum geomagnetic activity for a given number of sunspots and can be considered as the contribution of sunspot-related solar agents to geomagnetic activity. The equation of this line is $aa_R = aa_0 + b.R$ where $aa_0$ is the geomagnetic activity "floor" – a minimum value below which the geomagnetic activity cannot fall even in the absence of any sunspots, and *b* is the sensitivity of geomagnetic activity to the increasing number of sunspots, that is to the sunspot-related solar activity [Georgieva et al., 2012].

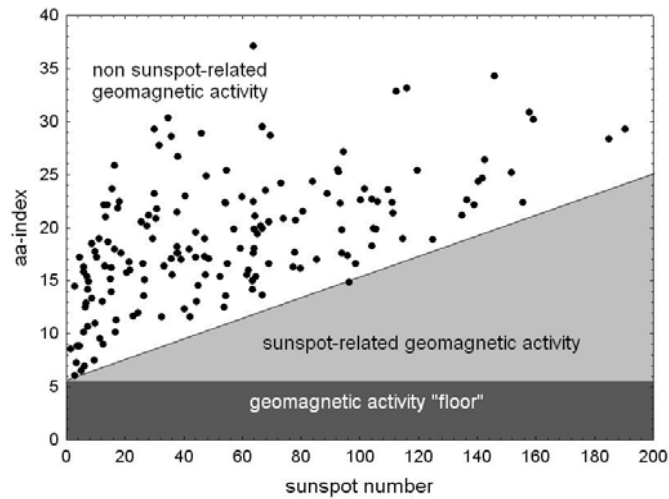

*Fig.1. Dependence of the aa-index of geomagnetic activity on the sunspot number: annual averages for the period 1868-2010.*

The coefficients $aa_0$ and *b* have cyclic long-term variations related to the quasi-secular solar (Gleissberg) cycle [Georgieva et al., 2012] – Fig.2. Cycle 24 is also included in Fig.2 though it is not yet complete, because as demonstrated by Kirov et al. [this issue], the coefficients $aa_0$ and *b* vary from cycle to cycle, but not within one cycle.

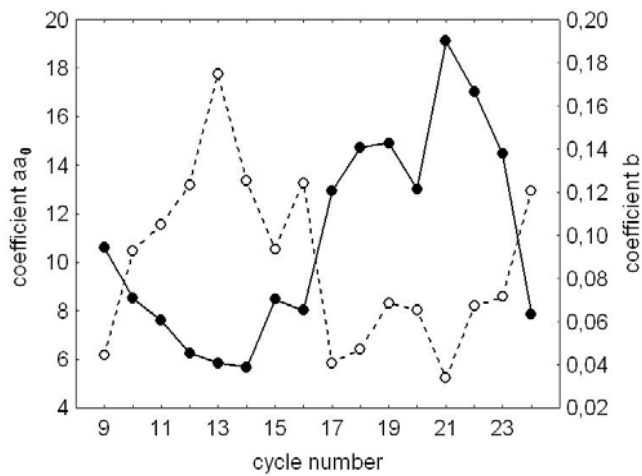

*Fig.2. Long term variations of the coefficients a (solid line) and b (dashed line).*

## 3. Correlations between the components of geomagnetic activity and the sunspot magnetic fields

It was found that the coefficient $aa_0$ (the geomagnetic activity floor, close to the geomagnetic activity in a sunspot cycle minimum) correlates well with the sunspot number in the cycle maximum: the higher the floor, the bigger the number of sunspots in sunspot maximum, with r = 0.792 with p=0.001 [Kirov et al., this issue]. On the other hand, Pevtsov et al. [2012] found a correlation between the sunspot field strength in sunspot minimum $B_{min}$ and the number of the sunspots in the cycle maximum: the stronger the magnetic field in sunspot minimum, the bigger the number of sunspots in the sunspot maximum. Fig.3 illustrates this correlation (the value of 57.7 for maximum of cycle 24 from http://sidc.oma.be/sunspot-data/ is still preliminary). Although the statistical sample is small, the correlation is very high and highly statistically significant: r = 0.997 with p<0.01.

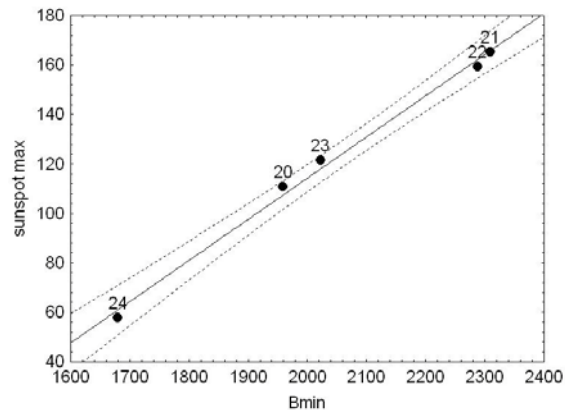

*Fig.3. Dependence of the sunspot number in solar cycle maximum on the sunspot magnetic field in the cycle minimum, with the 0.95 confidence limits (dotted lines).*

The correlations between the geomagnetic activity floor and the maximum sunspot number in the solar cycle, and between the sunspot magnetic field in sunspot minimum and the maximum sunspot number in the solar cycle, imply a correlation between the sunspot magnetic field in sunspot minimum $B_{min}$ and the geomagnetic activity floor $aa_0$. Indeed, Fig.4 demonstrates that this correlation is very high and highly statistically significant (r = 0.985 with p=0.002).

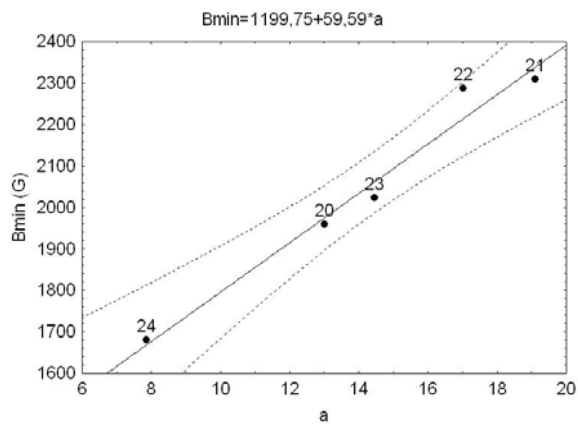

*Fig.4. Dependence of the geomagnetic activity floor on the sunspot magnetic field in the cycle minimum, with the 0.95 confidence limits (dotted lines).*

What can be the physical basis of this correlation? The geomagnetic activity floor $aa_0$ in cycle minimum is supposed to be determined by a constant baseline due to the slow solar wind from the solar equatorial streamer belt in which the Earth is immersed most of the time during all sunspot minima, plus a variable portion of additional geomagnetic activity originating from outside the streamer belt, which is different in different minima [Cliver and Ling, 2011]. Kirov et al. [2013, this issue] found that this additional geomagnetic activity is correlated to the occurrence frequency of moderate (with $aa$-index between 10 and 30) gradual commencement geomagnetic storms which are caused by high-speed solar wind from solar coronal holes. The coronal holes are unipolar areas of open flux whose properties are supposed to be determined by the properties and subsequent evolution of the emerging bipoles (sunspot pairs) because they are formed as a result of the decay of these sunspot pairs [Mackay and Yeates, 2012]. The correlation between the sunspot magnetic field and geomagnetic activity floor confirms that.

The sunspot-related geoeffective solar agents are the coronal mass ejections (CMEs) which, like sunspots, are manifestation of the solar toroidal field. The average geoeffectiveness of CMEs practically doesn't change in the solar cycle, so their varying contribution to the geomagnetic activity is caused by the variations in their number [Georgieva et al., 2006]. With increasing number of sunspots the number of CMEs also increases [Gopalswamy, 2010]. In the same time, with increasing number of sunspot the magnetic fields in sunspots increase. The correlation between the sunspot magnetic field and the number of CMEs is 0.57 with p=0.02 (Fig.5).

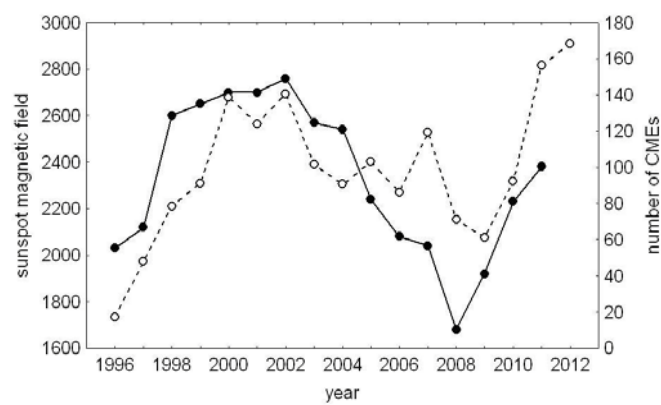

*Fig.5. Average yearly sunspot magnetic field (solid line) and number of CMEs (dashed line)*

Further, it was found [Georgieva, 2012] that the rate of increase of the sunspot magnetic field from sunspot minimum to sunspot maximum (grad$B$) is not constant but varies from cycle to cycle, and is strongly correlated (r=0.923 with p=0.025) to the coefficient $b$ the sensitivity of geomagnetic activity to increasing sunspot number – in the equation $aa_T = b*R$ (Fig.6). This correlation explains the correlation between $b$ and grad$B$: in cycles in which the sunspot magnetic field increases faster with increasing sunspot number, so does the number of geoeffective CMEs.

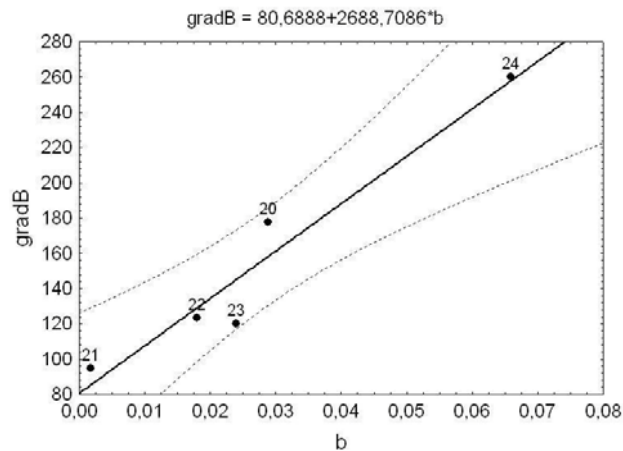

*Fig.6. Correlation between the rate of increase of the sunspot magnetic field from cycle minimum to maximum and the rate of increase of the geomagnetic activity with increasing sunspot number.*

**Reconstruction of the sunspot magnetic field from geomagnetic data**

The correlations between the geomagnetic activity and sunspot magnetic field parameters make it possible to estimate the long-term variations in the sunspot magnetic field. Using the regressions found, we can calculate $B_{min}$ and grad$B$. Further, knowing $B_{min}$, grad$B$ and the observed rise time of the sunspot cycle, we can also calculate $B_{max}$ – the sunspot magnetic field in cycle maximum. Fig.7 demonstrates the reconstruction (with linear interpolation between minima and maxima). The correlation between the calculated and measured values of the magnetic field in the cycle maxima is r=0.985 with p=0.02, and in the cycle maxima r= 0.895 with p=0.016. Note that, because the ascending part of the sunspot magnetic field cycle is assumed to coincide with the ascending part of the sunspot number cycle, so some discrepancies appear around the maxima of cycles in which the sunspot magnetic field maximum lags the sunspot number maximum (e.g. cycles 21 and 23).

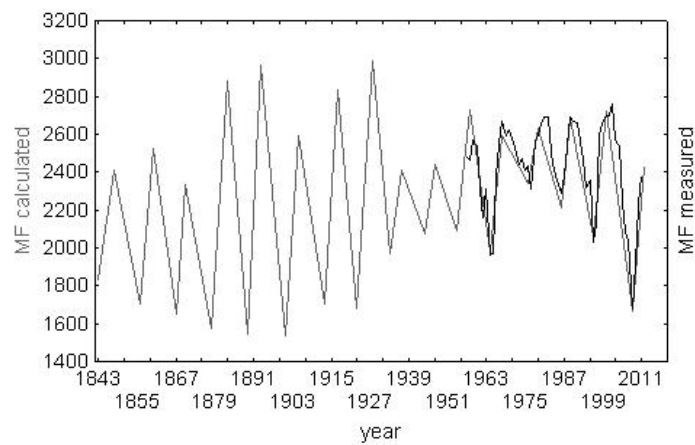

*Fig.7. Calculated (grey line) and measured values of the sunspot magnetic fields.*

Several features can be noted in the figure: The maximum sunspot number in a cycle is not correlated with the maximum sunspot magnetic field: the highest sunspot maximum magnetic fields are estimated to be in cycles 12-16 in which the sunspot activity was in its secular minimum. The sunspot maximum, however, does correlate with the sunspot magnetic field in sunspot minimum.

## A possible explanation of the sunspot magnetic field variations

Pevtsov et al. [2012] suggested that the sunspot cycle dependence of the sunspot magnetic field strength may be due to the different depths at which the sunspots originate. Javaraiah and Gokhale [1997] compared the "initial" rotation velocity (the rotation velocity when they are first seen) of sunspot groups of different age, size and magnetic field strength, and the radial rotation profile in the solar convection zone derived by helioseismology, and found that sunspots with stronger magnetic fields originate deeper in the convection zone. Therefore, the solar cycle variation of the sunspot magnetic field strength may indeed indicate a solar cycle variation of the depth at which sunspots originate. What about the cycle to cycle variations?

The solar toroidal magnetic field whose manifestations are the sunspots is generated in the lower part of the solar convection zone, in the equatorward branch of the large-scale solar meridional circulation. As shown in [Georgieva, 2012], the depth where the direction of the meridional circulation (and respectively, the depth below which the toroidal field is generated) has cycle to cycle variations and is related to the amplitude of the following sunspot maximum with higher sunspot maximum after a deeper reversal. A comparison of $B_{min}$ and the reversal depth of the meridional circulation calculated as described in [Georgieva, 2012] shows that the sunspot magnetic field in cycle minimum is stronger when the meridional circulation reverses direction deeper in the convection zone (Fig.8) – that is, when the sunspots originate deeper in the convection zone. The correlation between $B_{min}$ and the reversal depth is r=0.9. The statistical significance is difficult to estimate because of the small size of the sample but the dependence is obvious.

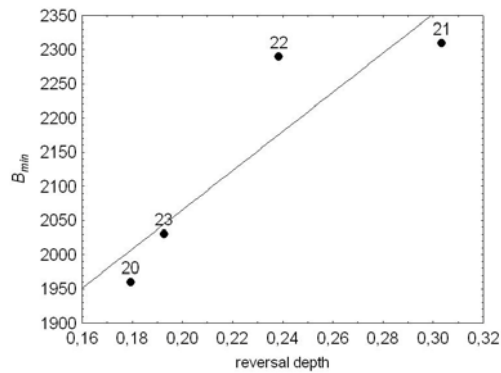

*Fig.8. Dependence of the sunspot magnetic field in cycle minimum on the reversal depth of the solar meridional circulation.*

## Summary and conclusion

Correlations are found between the cycle to cycle variations of the sunspot magnetic fields and the parameters of geomagnetic activity. The "floor" under which the geomagnetic activity cannot fall even in the absence of any sporadic or recurrent solar events is correlated to the sunspot magnetic field in the sunspot cycle minimum. The sensitivity of the geomagnetic activity to the increase in sunspot numbers in the course of the solar cycle vary depending on the rate of increase of the sunspot magnetic fields from sunspot minimum to sunspot maximum.

Based on these correlations, the long-term variations of sunspot magnetic fields in the cycles' minima and maxima can be estimated from geomagnetic data.

We have demonstrated that geomagnetic data can be used to estimate, which can help to better understand the operation of solar dynamo and solar influences on the Earth. These estimations reveal that the long-term variations of sunspot activity are due to long-term variations of the sunspot magnetic fields. The variations of the sunspot magnetic fields are possibly due to the variations of the depth where the sunspots originate. A relation is demonstrated between the reversal depth of the large-scale solar meridional circulation below which sunspots are formed and the sunspot magnetic field in sunspot minimum.

REFERENCES

Abdussamatov, H. I. On the Physical Relation between the Magnetic Field and the Brightness in the Sunspot Umbrae// Bulletin of the Astronomical Institute of Czechoslovakia Vol. 24, P.118-120, 1973.
Albregtsen F., Maltby P., New light on sunspot darkness and the solar cycle//Nature Vol. 274, P. 41-42, 1978.
Cliver E.W., Ling A.G. The Floor in the Solar Wind Magnetic Field Revisited//Solar Physics D Vol. 274, Issue 1-2, P. 285-301, 2011.
Feynman J. Geomagnetic and solar wind cycles, 1900-1975// J. Geophys. Res. Vol. 87, P.6153-6162 , 1982.
Georgieva K., Kirov B., Gavruseva E. Geoeffectiveness of different solar drivers, and long-term variations of the correlation between sunspot and geomagnetic activity// Physics and Chemistry of the Earth Vol. 31, Issue 1-3, P. 81-87, 2006.
Georgieva K., Kirov B., Koucká Knížová P., Mošna Z., Kouba D., Asenovska Y. Solar influences on atmospheric circulation//Journal of Atmospheric and Solar-Terrestrial Physics Vol. 90, P. 15-25, 2012.
Georgieva, K. Space Weather and Space Climate – What the look from the Earth tells us about the Sun//The environments of the Sun and the stars, Lecture Notes in Physics Vol. 857 Springer-Verlag Berlin Heidelberg, 253 p., 2013.
Gopalswamy N. Coronal mass ejections: a summary of recent results// 20th National Slovak Solar Physics Meeting, P. 108-130, 2010.
Hale G.E. On the Probable Existence of a Magnetic Field in Sun-Spots// Astrophysical Journal. Vol. 28, P.315-342, 1908.
Houtgast J., van Sluiters A. Statistical investigations concerning the magnetic fields of sunspots I// Bulletin of the Astronomical Institutes of the Netherlands, Vol. 10 P.325-333, 1948.
Javaraiah J., Gokhale M.H. Estimation of the depths of initial anchoring and the rising-rates of sunspot magnetic structures from rotation frequencies of sunspot groups//Astron. Astrophys. Vol. 327, P.795-799, 1997.
Kirov B., Obridko V.N., Georgieva K., Nepomnyashtaya E.V., Shelting B.D. Long-term variations of geomagnetic activity and their solar sources//This issue
Mackay D., Yeates A. The Sun's Global Photospheric and Coronal Magnetic Fields: Observations and Models//Living Reviews in Solar Physics, Vol. 9, No. 6, URL (cited on 10.01.2013): http://www.livingreviews.org/lrsp-2012-6
Nagovitsyn Yu.A., Pevtsov A.A., Livingston W.C., On a Possible Explanation of the Long-term Decrease in Sunspot Field Strength//Astrophys. J. Lett. Vol. 758 No 1, L20, 2012.
Penn M.J., Livingston W. Are sunspots different during this solar minimum?// Eos Vol. 90, P. 257-258, 2009.
Penn M.J., Livingston W. Long-term evolution of sunspot magnetic tields// arXiv:1009.0784v1 [astro-ph.SR], 2010.
Pevtsov A.A., Nagovitsyn Yu.A., Tlatov A.G., Rybak A.L. Long-term Trends in Sunspot Magnetic Fields// Astrophys. J. Lett. Vol. 742, L36-L39, 2011.
Ringnes TS., Jensen, E. On the relation between magnetic fields and areas of sunspots in the interval 1917-56// Astrophisica Norvegica Vol. 7, P.99-121, 1960.